Chapter 3

**The Creative Process of Cultural Evolution**

Liane Gabora
University of British Columbia

Correspondence should be addressed to:
Liane Gabora
Department of Psychology, University of British Columbia
Okanagan Campus, 1147 Research Road, Kelowna BC, Canada V1V 1V7
Email: liane.gabora@ubc.ca
Tel: 250-807-9849

**Contents**

## Introduction

"If I have seen further it is by standing on the shoulders of giants."
--Isaac Newton, 16[th] Century

Even this saying itself is a variant of a similar statement attributed to Bernard of Chartres in the 12th Century, and inspired the title for a book by Steven Hawking and an album by Oasis. Creative ideas beget other creative ideas and, as a result, modifications accumulate, and we see an overall increase in the complexity of cultural novelty over time, a phenomenon sometimes referred to as the *ratchet effect* (Tomasello, Kruger, & Ratner, 1993). Although we may never meet the people or objects that creatively influence us, by assimilating what we encounter around us and bringing to bear our own insights and perspectives, we all contribute in our own way, however small, to a second evolutionary process – the evolution of culture.

This chapter explores how we can better understand culture by understanding the creative processes that fuel it, and better understand creativity by examining it from its cultural context. First, we look at some theoretical frameworks for how culture evolves and what these frameworks imply for the role of creativity. Then we will see how questions about the relationship between creativity and cultural evolution have been addressed using an agent-based model. We will also discuss studies of how creative outputs are influenced, in perhaps unexpected ways, by other ideas and individuals, and how individual creative styles "peek through" cultural outputs in different domains.

### A Scientific Framework for Cultural Evolution

Darwin vastly enhanced our understanding of the organismal world by integrating scattered biological knowledge into an evolutionary framework, enabling us to see how species fit together in a unified "tree of life". His feat even improved our ability to make predictions about what kinds of underlying mechanisms were at work and what traits or species we might expect in particular environments. Since art, technology, languages, and customs change over time in a manner seemingly reminiscent of biological evolution, it seemed reasonable to view culture as a second evolutionary process. This section gives a brief overview of two theories of cultural evolution and discusses what these theories imply for our understanding of creativity.

#### Cultural Evolution as a Selectionist Process

Biological evolutionary processes can be seen as consisting of two components: the *generation* of new variants, and the differential survival or *selection* of some of those variants, such that they live long enough to produce offspring. Darwin's explanation focused not on the generation of variants but on the selection of some fraction of them, which is why it is sometimes referred to as a *selectionist* theory. He posited that biological change is due to the effect of differential selection on the distribution of heritable variation in a population; those with adaptive traits live longer, have more offspring, and are 'selected' for, and therefore, their traits proliferate in future generations. Thus, a selectionist process works through competitive exclusion amongst *existing* variants; it is not a mechanism that affects how variants are generated in the first place. In fact, it assumes that variants are generated randomly, for to the extent that variation is not randomly generated, the distribution of variants reflects whatever was biasing the generation away from random in the first place, not survival of the fittest. Notice also that the theory operates on the timescale of generations, as it requires at a minimum an entire generation for change to occur.

The reason a selectionist explanation is applicable in biology is that in biological evolution there are two kinds of traits: (1) *inherited traits* (e.g., blood type or eye color), which are transmitted vertically from parent to offspring by way of the genes, and (2) *acquired traits* (e.g., a tattoo, or knowledge of someone's name), which are obtained during an individual's lifetime, and which are sometimes called *epigenetic* because they are not transmitted vertically through the genes. Since acquired traits are not passed down (e.g., you do not inherit your mother's tattoo), the fast intra-generational transmission of acquired traits does not drown out the slow, intergenerational transmission of inherited traits. In other words, a Darwinian explanation works when *acquired* change is negligible relative to *inherited* change; otherwise the first, which can operate over minutes, overwhelms the second, which requires generations.

Attempts to develop a scientific framework for cultural evolution initially framed it as a selectionist process (Aunger, 2000; Boyd & Richerson, 1985; Gabora, 1996). However, this was difficult to reconcile with the highly cooperative nature of human societies and, when examined closely, the selectionist theory is incompatible with some basic facts about how culture evolves (Fracchia & Lewontin, 1999; Gabora, 2004, 2011, 2013, Gabora & Kauffman, in press; Temkin & Eldredge, 2007). In cultural evolution there is no mechanism for discarding acquired change (e.g., once one cup had a handle, all cups could have handles). Therefore, acquired change accumulates orders of magnitude faster than, and overwhelms, change due to the mechanism Darwin proposed: differential replication of variants in response to selection. Moreover, to the extent that the *generation* of novelty deviates from random chance, change is due to whatever is causing that deviation in the first place rather than to *selection* of fitter variants. In short, cultural change is acquired, not inherited, and it is generated not randomly, but through strategy and intuition. Therefore, a Darwinian theory of cultural evolution has been found to be inappropriate (Gabora, 2004, 2008a, 2011, 2013; Gabora & Kauffman, in press).

## Cultural Evolution as a Communal Exchange Process

This analysis prohibits a selectionist, but not an evolutionary, framework for culture. Even in the biological realm we are only just starting to appreciate the key role played by non-Darwinian epigenetic processes (Kauffman, 1993; Koonin, 2009; Koonin & Wolf, 2012; Lynch, 2007; Woese, 2002; Woese, Goldenfeld, & 2009). Research on the origin of life suggests that early life consisted of autocatalytic protocells that evolved through a non-Darwinian process of *communal exchange*, and that natural selection emerged later from this more haphazard, ancestral evolutionary process (e.g., Gabora, 2006; Segre, 2000; Vetsigian, Woese, & Goldenfeld, 2006; Wächtershäuser, 1997). Communal exchange is more haphazard than selection and does not require a self-assembly code (such as the genetic code). What it requires is structure that is (1) *self-organizing:* its components generate new components through their interactions, (2) *self-replicating:* through duplication of components it can reconstitute an entity like itself, and (3) *interactive:* entities exchange components. **Table 1** provides a summary of the similarities and differences between natural selection and communal exchange. Both have mechanisms for preserving continuity and for introducing novelty. However, whereas natural selection is a high fidelity Darwinian process and the structure that self-replicates is DNA-based self-assembly instructions, communal exchange is a low fidelity Lamarckian process, and the structure that replicates is an autopoietic network. Only communal exchange allows transmission of acquired traits. Communal exchange is the proposed mechanism by which early life evolved, as well as the mechanism by which culture evolves, and some aspects of present day life, such as horizontal gene transfer. A schematic illustration of both evolution through a selectionist process and evolution through communal exchange is provided in **Figure 1**.

**Table 1.** Similarities and differences between two evolutionary frameworks: natural selection and communal exchange. From (Gabora, 2013).

|  | **Natural Selection** | **Communal Exchange** |
|---|---|---|
| Unit of self-replication | DNA | Autopoietic network |
| Mechanism for preserving continuity | DNA replication; proofreading enzymes, etc. | Retention of horizontally transmitted information |
| Mechanism for generating novelty | Mutation; recombination; replication errors; pseudo-genes | Faulty duplication of autopoietic structure; transmission errors; innovation |
| Self-assembly code | Yes | No |
| High fidelity | Yes | No |
| Transmission of acquired traits | No | Yes |
| Type of process | Darwinian | Lamarckian (by some standards) |
| Evolutionary processes it seeks to explain | Biological | Early life; horizontal gene transfer (HGT); cultural |

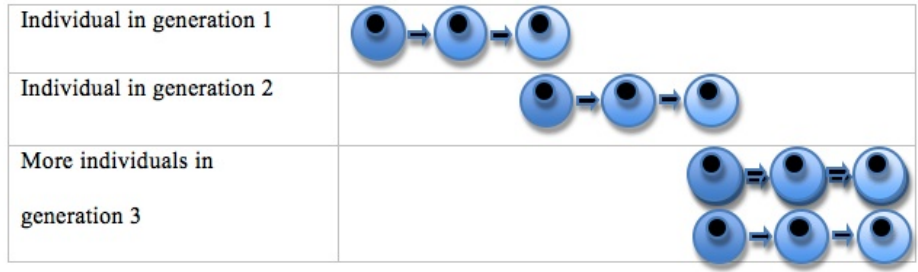

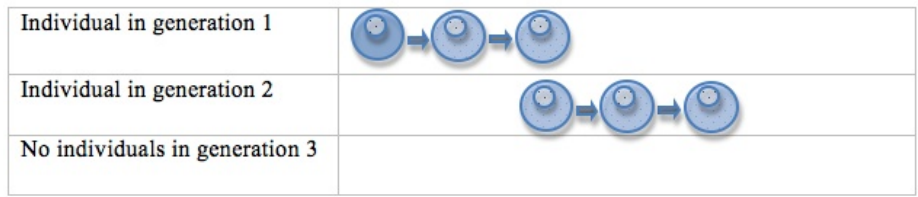

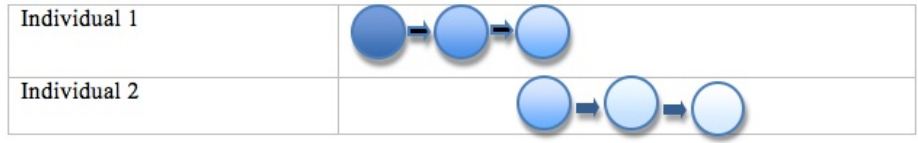

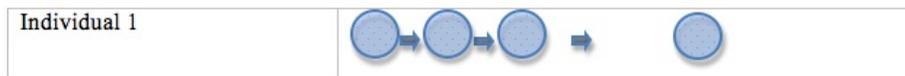

Figure 1. Top: Schematic depiction of evolution through natural selection. Each individual has a genotype, represented by a small circle, and a phenotype, represented by a larger circle encompassing the small circle. Phenotypic change acquired over a lifetime, indicated by increasingly lighter color, is not passed on, as indicated by the fact that in lineage A the individual born in generation 2 did not inherit the lighter color the parent acquired by the time it had offspring. Lineage A outcompetes lineage B as indicated by the fact that by generation 3 there are no individuals left in lineage B. Evolution relies on *competition amongst* individuals rather than *transformation of* individuals. Bottom: Evolution through *communal exchange* involves the duplication and transformation of variants. There is no distinction between phenotype and genotype, and acquired characteristics are transmitted, as shown by how in lineage A, Individual 2 has the light color its 'parent' acquired over its lifetime. Death does not have the same finality since, given the right conditions, an inert individual could potentially reconstitute itself, as in Lineage B. Therefore the term 'generations' is not meaningful. Acquired change is transmitted. Communal exchange relies more on *transformation of* individuals than *competition amongst* individuals. Biological evolution involves both communal exchange and natural selection. With respect to cultural evolution the matter is still a subject of discussion but many, including the author, believe it evolves solely through communal exchange. From (Gabora, 2013).

If even biological life originally evolved through communal exchange, it seems reasonable that cultural evolution would get established through this simpler, more haphazard process (Gabora, 2000a, 2013). Adults share ideas with children such that eventually they develop their own self-organized network of understandings, at which point they can adapt ideas to their own needs and perspectives and thereby contribute creatively to culture. We can refer to such a web of understandings as a *worldview*. A worldview is self-regenerating: an adult shares concepts, ideas, attitudes, stories, and experiences with children (and other adults), influencing little by little the formation of other worldviews. Thus, through interactions amongst its parts, a worldview not only responds to perturbations, but reconstitutes itself. Each worldview takes form through the influence of many others, though some, such as those of relatives and teachers, may predominate. Children expose fragments of what was originally the adult's worldview to different experiences and different bodily constraints, thereby forging unique internal models of the relation of self to world.

Thus the human mind dynamically reconfigures itself to achieve a more stable configuration, and acts in ways that cause such reconfigurations to proliferate; it possesses the necessary structure for evolution through communal exchange, as illustrated schematically in **Figure 2**. The more information a mind encodes, the greater the variety of ways this knowledge base can be reconfigured. Its configuration is revealed through behavioural regularities in how it expresses itself and responds to situations, and in the creative outputs it generates. Note that although this process does not involve selection in its technical sense (i.e., change due to the effect of differential selection on the distribution of heritable variation across a population), selection as the term is used in everyday parlance may play a role (i.e., individuals may be selective about which aspects of their worldviews they share or assimilate).

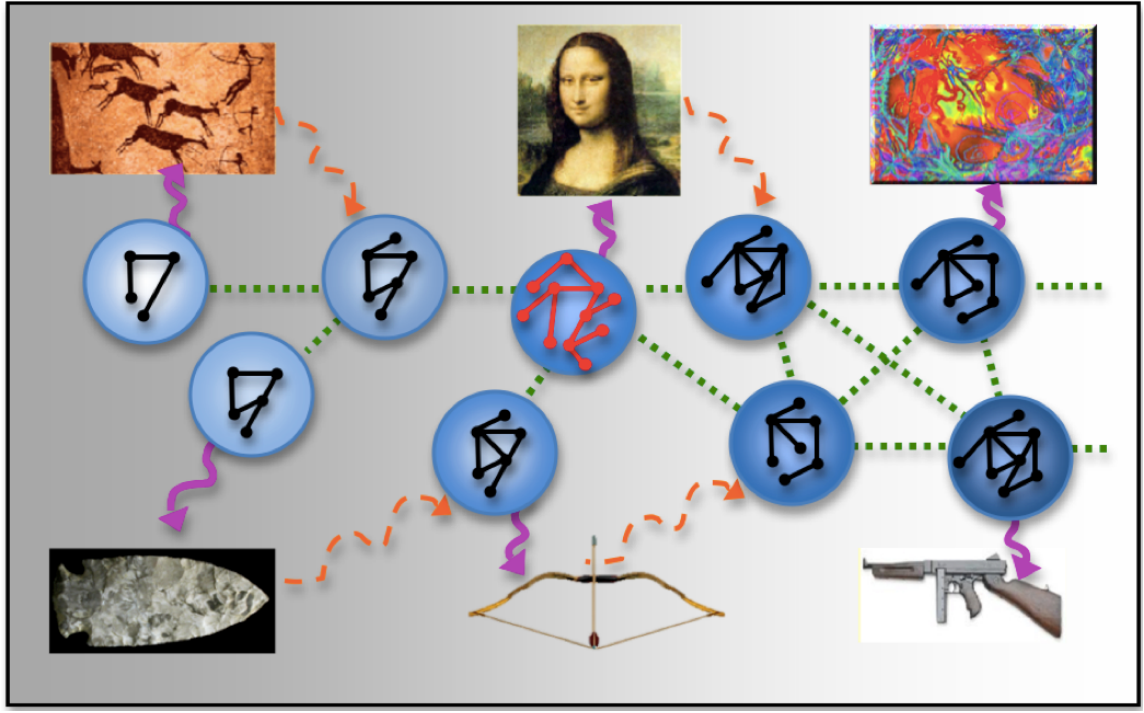

Figure 2. Schematic depiction of how worldviews evolve, not through *survival of the fittest* but *transformation of all. Worldviews* transform as a consequence of psychological entropy-reducing restructuring and communal exchange. Individuals are represented by spheres and their internal models of the world, or *worldviews*, are represented by networks within the spheres. Patterns of social transmission are indicated by dashed green lines. Creative contributions to culture are indicated by wavy purple arrows from creator to artifact. Learning through exposure to artifacts is indicated by wavy orange arrows. Worldviews and patterns of social transmission tend to become more complex over time. Individuals such as the one with the red network are more compelled than others to reframe what they learn in their own terms, potentially resulting in a more unique or nuanced worldview. Such 'self-made' individuals are more likely to have an effect on culture (e.g., through the creation of artifacts), which may influence the formation of new worldviews long after they have died. This is indicated by the diffusion of segments from self-made individuals to other individuals, either indirectly by way of exposure to the Mona Lisa, or directly by way of exposure to its creator. From (Gabora, 2013).

For minds to evolve through communal exchange they must be organized such that, for any given concept or idea, there exists *some* pathway (e.g., a chain of associations, or deductive reasoning) by which it could potentially interact with and modify any other given concept or idea. The concepts and ideas must form an integrated whole, i.e., they must be able to interact with and modify others not just in the same local cluster (as in comparing one dress to another), but across clusters (as in cross-domain analogy). Thus, a big question is: how did the human mind acquire this kind of structure?

*Evolution of Capacity for Chaining and Redescription*

Let us first consider how the mind acquired the capacity to modify thoughts and ideas by thinking about them in the context of other thoughts and ideas that are similar, that is, in the same local cluster. Merlin Donald (1991) suggested that the enlarged cranial capacity of our *Homo erectus* ancestors 1.7 million years ago enabled them to voluntarily retrieve and modify memories independent of environmental cues (sometimes referred to as 'autocuing'), a capacity he referred to as *self-triggered recall and rehearsal,* and which ushered forth a transition to a new mode of cognitive functioning. Thus, while *Homo habilis* was limited to the "here and now", *Homo erectus* could *chain* memories, thoughts, and actions into more complex ones, and progressively modify them, thereby gaining new perspectives on past or possible events, and even mime or re-enact them for others. The notion of self-triggered recall bears some resemblance to Hauser et al.'s (2002) idea that what distinguishes human cognition from that of other species is the capacity for recursion, and to Penn, Penn, Holyoak, and Povinelli's (2008) concept of *relational reinterpretation*, the ability to reinterpret higher order relations between perceptual relations.

Donald's proposal has been shown to be consistent with the structure and dynamics of associative memory (Gabora, 2000b, 2010a). Neurons are sensitive to primitive stimulus attributes or 'microfeatures', such as sounds of a particular pitch or lines of a particular orientation. Experiences encoded in memory are *distributed* across cell assemblies of neurons, and each neuron participates in the encoding of many experiences. Memory is also *content-addressable:* similar stimuli activate overlapping distributions of neurons. With larger brains, experiences could be encoded in more detail, enabling a transition from coarse-grained to fine-grained memory. Fine-grained memory enabled concepts and ideas to be encoded in more detail, that is, there were more ways in which distributed sets of microfeatures could overlap. Greater overlap enabled more routes for self-triggered recall, and paved the way for streams of abstract thought. Ideas could now be reprocessed until they fit together with cognitive structures already in place, allowing for the emergence of local clusters of mutually consistent ideas, and thus for a more coherent internal model of the world, or worldview (Gabora, 1999).

*Evolution of Capacity for Cross-Domain Thinking*

At this point it was possible to think about an idea in relation to other closely related ideas and thereby forge clusters of mutually consistent ideas, which allowed for a narrow kind of creativity, limited to minor adaptations of existing ideas. But the mind was not integrated, nor truly creative, until it could forge connections between seemingly disparate ideas as in the formation of analogies. How did this come about?

One proposal is that it was due to the onset in the Middle/Upper Paleolithic of *contextual focus* (CF): the ability to shift between different modes of thought—an explicit *analytic mode* conducive to logical problem solving, and an implicit *associative mode* conducive to insight and breaking out of a rut (Gabora, 2003). While dual processing theories generally attribute abstract, hypothetical thinking solely to the more recently evolved "deliberate" mode (e.g., Evans, 2003), according to the CF hypothesis it is possible in either mode but it will differ character in the two modes (flights of fancy versus logically constructed arguments) (Sowden, Pringle, & Gabora, 2014). CF thus paved the way for integration of different domains of knowledge (Mithen, 1998).

It has been proposed that CF was made possible by mutation of the FOXP2 gene, which is known to have undergone human-specific mutations in the Paleolithic era (Chrusch & Gabora, 2014; Gabora & Smith, submitted). FOXP2, once thought to be the "language gene", is not uniquely associated with language. The idea is that, in its modern form, FOXP2 enabled fine-tuning of the neurological mechanisms underlying the capacity to shift between processing modes by varying the size of the activated region of memory.

*Implications of Cultural Evolution Framework for Creativity*

The communal exchange theory of cultural evolution suggests a theory of creativity, sometimes referred to as honing theory, according to which peoples' uniquely structured webs of understanding, or *worldviews*, are the basic units of cultural evolution. It is through the 'honing' of creative ideas that worldviews transform and evolve. In other words, the creative process reflects the natural tendency of a worldview to seek a state of dynamic equilibrium by exploring perspectives and associations until the worldview achieves a more stable state than it started with. A creative outcome (e.g., a painting) can be an *external* manifestation of *internal* cognitive restructuring brought about through immersion in a creative task. The creator may see and feel the world differently afterward, which is why creativity can be transformative, and why expressive art therapies are gaining prominence. Not all creative outputs involve extensive cognitive restructuring; some are minor variations on a theme, and others outright imitations, which is why a creative work cannot be fully understood outside of its cultural context.

## Computationally Modeling the Evolution of Cultural Novelty

It is difficult to experimentally test hypotheses about how the creative abilities underlying cultural transitions evolved. Agent-based modeling is a computational methodology in which artificial agents can be used to represent interacting individuals. It enables us to address questions about the workings of collectives such as societies. It is particularly valuable for answering questions of this sort which lie at the interface between anthropology and psychology, owing both to (1) the difficulty of experimentally manipulating a variable, such as the average amount by which one invention differs from its predecessor and observing its impact on cumulative culture over time, and (2) the sparseness of the pre-modern archaeological record. Although methods for analyzing these remains are becoming increasingly sophisticated, they cannot always distinguish amongst different theories.

EVOC (for EVOlution of Culture) is a computational modeling of cultural evolution that consists of neural network based agents that invent new actions and imitate actions performed by neighbors (Gabora, 1995, 2008b). The assemblage of ideas changes over time not because some replicate at the expense of others, as in natural selection, but through inventive and social processes. Agents can learn generalizations concerning what kinds of actions are useful, or have a high "fitness", with respect to a particular goal, and use this acquired knowledge to modify ideas for actions before transmitting them to other agents. A model such as EVOC is a vast simplification, and results obtained with it may or may not have direct bearing on complex human societies, but it allows us to vary one parameter while holding others constant and thereby test hypotheses that could otherwise not be tested. It provides new ways of thinking about and understand what is going on.

EVOC exhibits typical evolutionary patterns, such as (1) a cumulative increase in the fitness and complexity of cultural outputs over time, and (2) an increase in diversity as the space of possibilities is explored followed by a decrease as agents converge on the fittest possibilities, as illustrated in **Figure 3**. EVOC has been used to model how the mean fitness and diversity of cultural elements is affected by factors such as population size and density, and borders that affect transmission between populations, as well as the questions reported here pertaining to creativity.

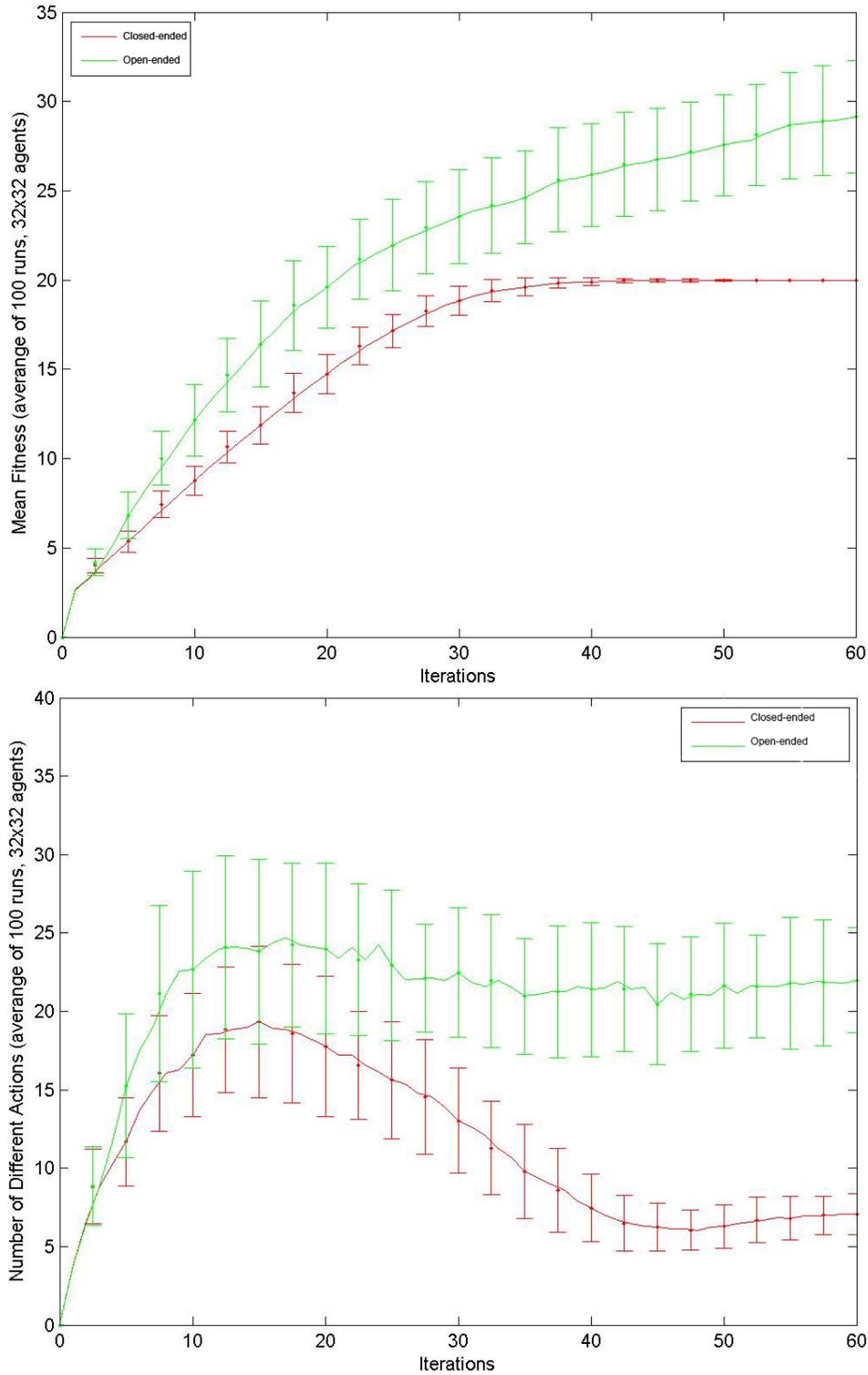

Figure 3. The typical increase in fitness of cultural outputs over time (top) and increase in diversity as the space of possibilities is being explored followed by a decline as the society converges on the fittest (bottom). These graphs also demonstrate the effect on fitness and diversity an open-ended space of possibilities.

**Modeling Hypothesized Cognitive Breakthroughs Underlying Cultural Transitions**

      Recall Donald's hypothesis that cultural evolution was made possible by the onset of the capacity for one thought to trigger another, leading to the chaining and progressive modification of thoughts and actions (Donald, 1991). This was tested in EVOC by comparing runs in which agents were limited to single-step actions to runs in which they could chain ideas together to generate multi-step actions (Gabora, Chia, & Firouzi, 2013). As illustrated in **Figure 4**, chaining increased the mean fitness and diversity of cultural outputs across the artificial society (Gabora, Chia, & Firouzi, 2013). While chaining and no-chaining runs both converged on optimal actions, without chaining this set was static, but with chaining it was in constant flux as ever-fitter actions were found. While without chaining there was a ceiling on mean fitness of actions, with chaining there was no such ceiling, and chaining also enhanced the effectiveness of the ability to learn trends. These findings support the hypothesis that the ability to chain ideas together can transform a culturally static society into one characterized by open-ended novelty.

      The hypothesis that the onset of contextual focus (CF) brought about a second cognitive transition underlying the human capacity to evolve complex culture was also tested with EVOC (Gabora, Chia, & Firouzi, 2013). When the fitness of an agent's outputs was low it temporarily shifted to a more divergent mode by increasing $\alpha$: the degree to which a newly invented idea deviates from the idea on which it was based. As illustrated in **Figure 4**, both mean fitness of actions across the society increased with CF, as hypothesized, and CF was particularly effective when the fitness function changed, which supports its hypothesized utility in breaking out of a rut and adapting to new or changing environments.

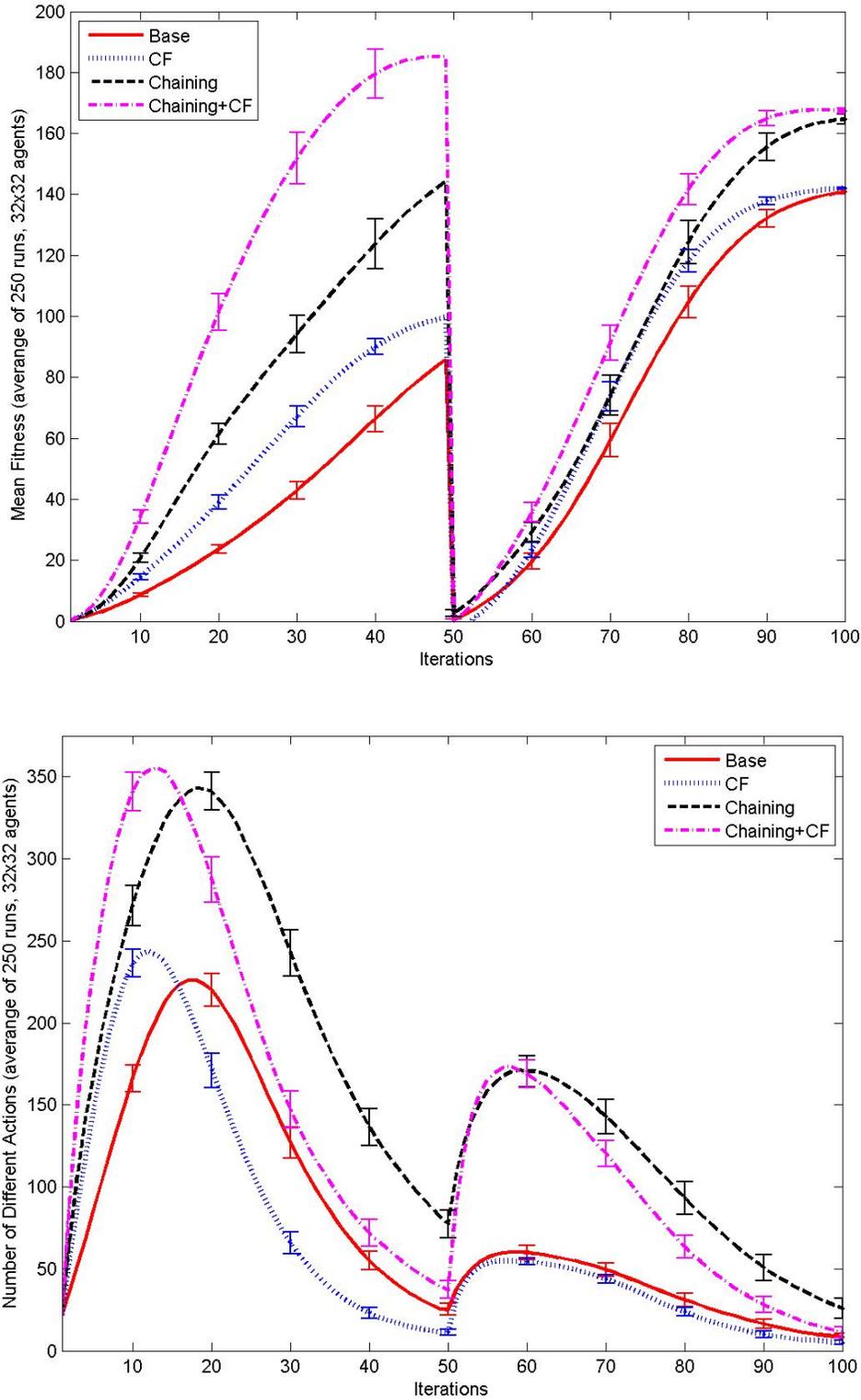

Figure 4. Mean fitness (top) and diversity (bottom) of cultural outputs across the artificial EVOC society with both chaining and CF, chaining only, CF only, and neither chaining nor CF. Data are means of 500 runs. (From Gabora & Smith, submitted.)

These findings show how a cultural evolutionary framework can provide a valuable perspective on creativity. Note however that although chaining made the variety of novel outputs open-ended, and this became even more pronounced with CF, these novel outputs were nonetheless predictable. Chaining and CF did not open up new cultural niches in the sense that, for example, the invention of cars created niches for the invention of things like seatbelts and stoplights. EVOC in its current form could not solve *insight problems*, which require restructuring the solution space. Nonetheless it is sufficient to illustrate the effectiveness of chaining and CF. Building on a related research program in concept combination (e.g., Aerts, Gabora, & Sozzo, 2013), models of concepts provide further support, showing that CF is conducive to making creative connections by placing concepts in new contexts (Gabora & Aerts 2009; Gabora & Kitto, 2012; Veloz, Gabora, Eyjolfson, & Aerts, 2011).

**Impact of Creative Leadership on Cultural Evolution**

Throughout history there have been leaders who were imitated more frequently than the common person. An interesting question is: what is the impact of leadership on creativity and the evolution of culture?

This question was also investigated in EVOC using the *broadcasting* function. Broadcasting allows the action of a leader to be visible to not just immediate neighbors, but all agents, thereby simulating the effects of media such as public performances, television, or the Internet, on patterns of cultural change. When broadcasting is turned on, each agent adds the broadcaster as a possible source of actions it can imitate. A particular agent can be chosen as the broadcaster before the run, or the broadcaster can be chosen at random, or the user can specify that the agent with the fittest action is the broadcaster. Broadcasting can be intermittent or continuous throughout the duration of a run.

Broadcasting produced a modest increase in the fitness of actions, but it accelerates convergence on optimal actions, thereby consistently reducing diversity. This can be seen in **Figure 5** by comparing column 1 with no leader to column 2 with one leader. Here we see the previously mentioned increase in diversity as the space of possibilities is explored followed by a decrease as agents converge on the fittest possibilities. The total number of different actions after 20 iterations decreases from eight to five when a leader is introduced, and the percentage of agents executing the most popular action increases from 41\% to 84\%. Thus leadership accentuates the normal plummet in diversity.

**Figure 5** also shows the impact of a dictatorial style of leadership (one broadcaster) versus a more distributed style of leadership (multiple leaders). As we go from one leader to five leaders the total number of different actions after 20 iterations increases from five to nine, and the percentage of agents executing the most popular action decreases from 84\% to 31\%. Thus, the leadership induced decrease in diversity is mitigated by a more distributed leadership.

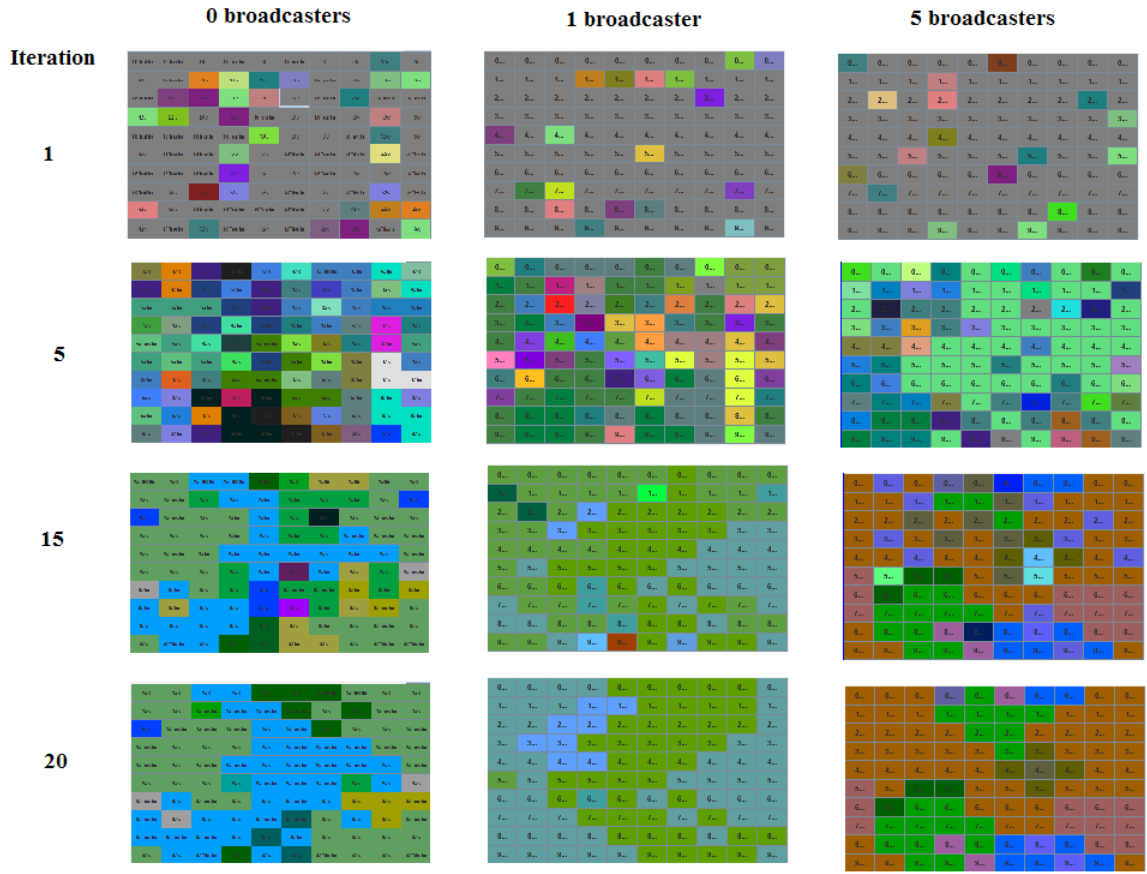

Figure 5. Diversity of actions over a run with 0, 1, and 5 broadcasters. Different actions are represented by differently coloured cells. In all cases there is an increase followed by a decrease in diversity over time (moving down any column from the first iteration at the top to the 20th at the bottom) but this becomes less pronounced with additional leaders. In these experiments broadcasters were chosen at random in every iteration, and when there were multiple broadcasters available to imitate, agents selected the broadcaster whose action was most similar to their own. (Adapted from Gabora, 2008b.)

The effectiveness of creative versus uncreative styles of leadership was also investigated. Creative leadership increased the mean fitness of cultural outputs only when non-leaders were relatively uncreative, and increased the diversity of outputs only early in a run during initial exploration of the space of possibilities (Leijnen & Gabora, 2010).

**Balancing Creativity with Continuity**

Cultural evolution, like any evolutionary process, requires a balance of change and continuity. While creative individuals generate the novelty that fuels cultural evolution, absorption in their creative process impedes the diffusion of proven solutions, effectively rupturing the fabric of society. Thus, it was hypothesized that a society in which creative (novelty injecting) individuals are interposed with imitating (continuity maintaining) individuals ensures both that new ideas come about and that, if effective, they are not easily lost by society as a whole. This hypothesis was tested in EVOC in a set of three experiments.

*Varying the Ratio of Inventing to Imitating*

        To investigate the optimal ratio of inventing to imitating, the invention-to-imitation ratio was systematically varied from 0 to 1. When agents never invented, there was nothing to imitate, and there was no cultural evolution at all. If the ratio of invention to imitation was even marginally greater than 0, not only was cumulative cultural evolution possible, but eventually all agents converged on optimal cultural outputs. When all agents always invented and never imitated, the mean fitness of cultural outputs was also sub-optimal because fit ideas were not dispersing through society. The society as a whole performed optimally when there was a mixture of inventing and imitating, with the optimal ratio of the two being approximately 1:1, with the exact value depending on the difficulty of the fitness function; for example, with the difficult fitness function shown in **Figure 6**, it was significantly lower than 1:1. This showed that, as in biological evolution, culture evolves most efficiently when the novelty-generating process of creativity is tempered with the continuity fostering process of imitation.

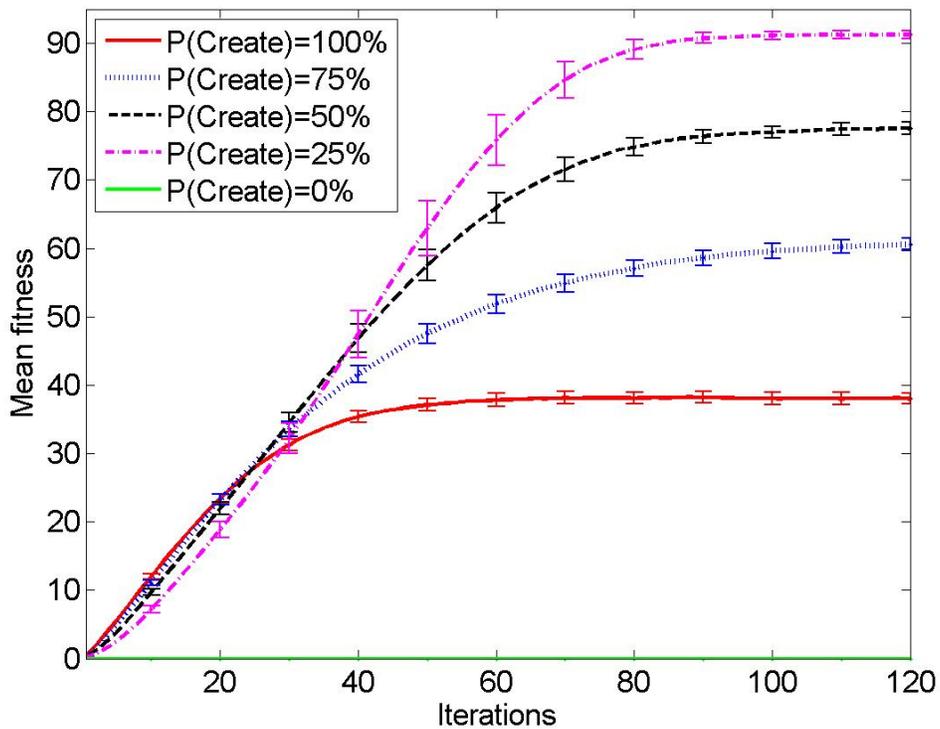

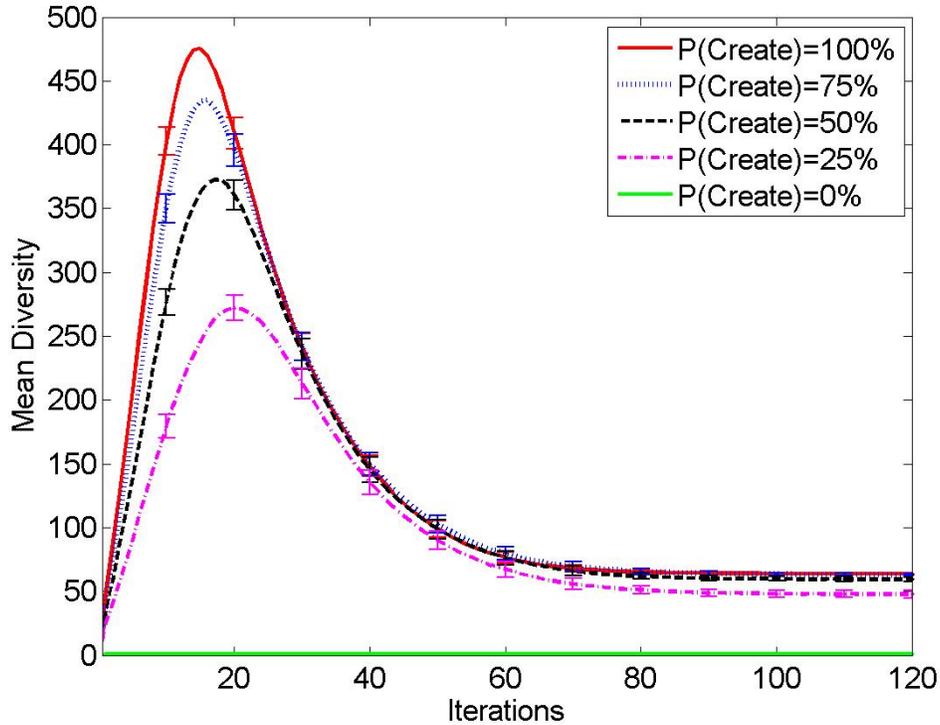

Figure 6. Fitness (top) and diversity (bottom) of cultural outputs with different ratios of inventing to imitating.

*Varying the Ratio of Creators to Conformers*

  The finding that very high levels of creativity can be detrimental for society led to the hypothesis that there is an adaptive value to society's ambivalent attitude toward creativity; society as a whole may benefit from a distinction between the conventional workforce and what has been called a 'creative class' (Florida, 2002). This was investigated by introducing two types of agents: *conformers* that only obtained new actions by imitating and *creators* that obtained new actions either by inventing or imitating (Gabora & Firouzi, 2012). A given agent was either a creator or an imitator throughout the entire run, and whether a given creator invented or imitated in a given iteration fluctuated stochastically. We could systematically vary *C*, the proportion of creators to imitators in the society, and *p*, how creative the creators were. As illustrated in **Figure 7**, we observed a tradeoff between *C* and *p* thus providing further evidence that society as a whole functions optimally when creativity is tempered with continuity.

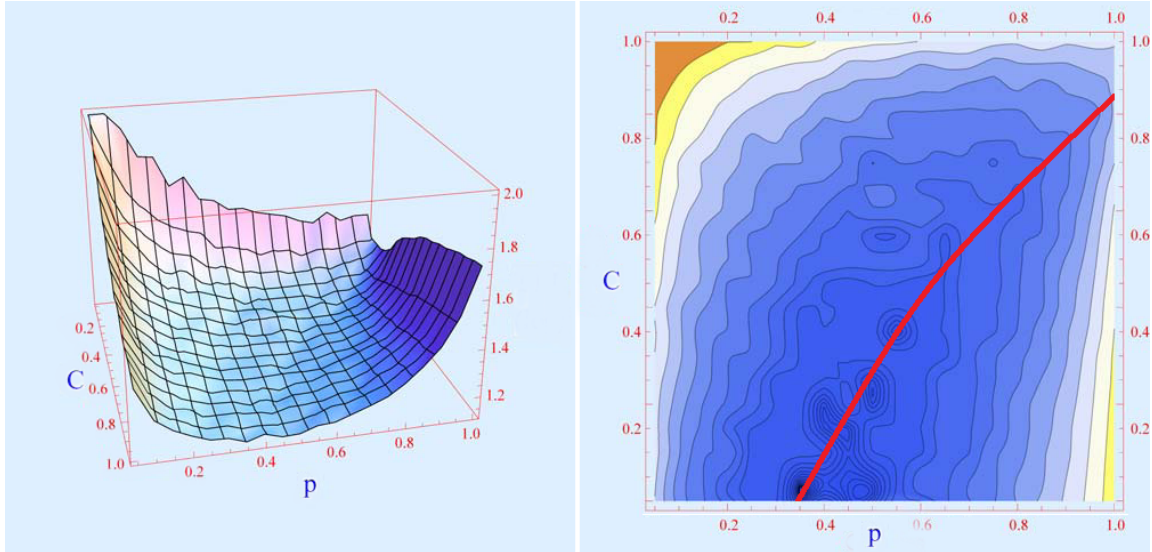

Figure 7. The effect of varying the percentage of creators, *C*, and how creative they are, *p*, on the mean fitness of ideas in EVOC. 3D graph (left) and contour plot (right) for the average mean fitness for different values of *C* and *p* using discounting to ensure that the present value of any given benefit with respect to idea fitness diminishes as a function of elapsed time before that benefit is realized. The z-axis is reversed to obtain an unobstructed view of surface; therefore, lower values indicate higher mean fitness. The red line in the contour plot shows the position of a clear ridge in fitness landscape indicating optimal values of *C* and *p* that are sub-maximal for most {*C, p*} settings, *i.e.,* a tradeoff between how many creators there are and how creative they should be. The results suggest that excess creativity at the individual level can be detrimental at the level of the society because creators invest in unproven ideas at the expense of propagating proven ideas. The same pattern of results was obtained analyzing just one point in time and using a different discounting method (not shown). (Adapted from Gabora, & Firouzi, 2012.)

*Social Regulation of How Creative People Are*

We then hypothesized that society as a whole might perform even better with the ability to adjust creativity in accordance with their perceived creative success, through mechanisms such as selective ostracization of deviant behavior unless accompanied by the generation of valuable cultural novelty, and encouraging of successful creators. A first step in investigating this was to determine whether it is algorithmically possible to increase the mean fitness of ideas in a society by enabling them to self-regulate how creative they are. To test the hypothesis that the mean fitness of cultural outputs across society increases faster with social regulation (SR) than without it, we increased the relative frequency of invention for agents that generated superior ideas, and decreased it for agents that generated inferior ideas (Gabora & Tseng, 2014). Thus when SR was on, if relative fitness was high the agent invented more, and if it was low the agent imitated more. *p(C)* was initialized at 0.5 for both SR and non-SR societies.

When social regulation was introduced into the artificial society, the mean fitness of the cultural outputs was higher, as shown in **Figure 8**. The typical pattern was observed with respect to diversity, or number of different ideas: an increase in diversity as the space of possibilities is explored followed by a decrease as agents converge on fit actions. However, this pattern occurred earlier, and was more pronounced, in societies with SR than societies without it.

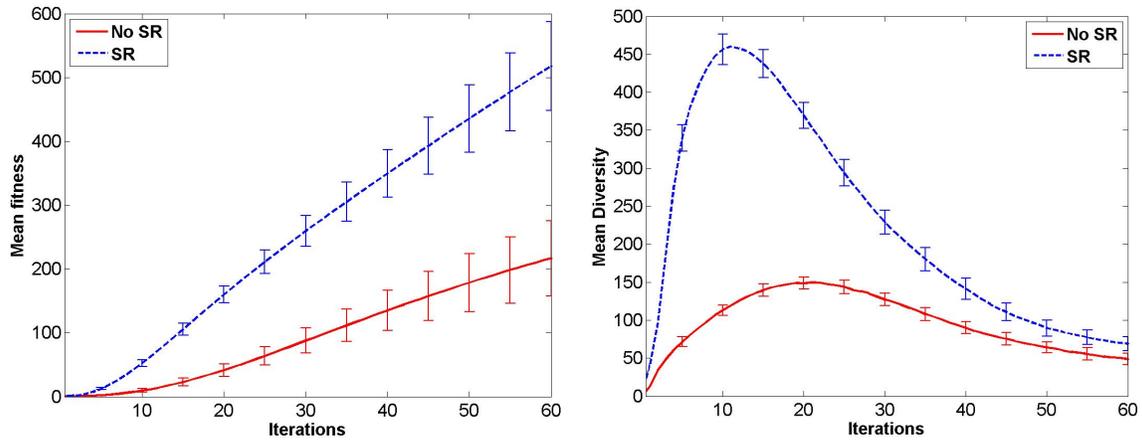

Figure 8. Mean fitness (left) and diversity (right) of cultural outputs across all agents over the duration of the run with and without social regulation. Data are averages across 250 runs. (From Gabora & Tseng, 2014).

As illustrated in **Figure 9**, societies with SR ended up separating into two distinct groups: one that primarily invented, and one that primarily imitated. Thus, the observed increase in fitness can indeed be attributed to increasingly pronounced individual differences in their degree of creative expression over the course of a run. Agents that generated superior cultural outputs had more opportunity to do so, while agents that generated inferior cultural outputs became more likely to propagate proven effective ideas rather than reinvent the wheel.

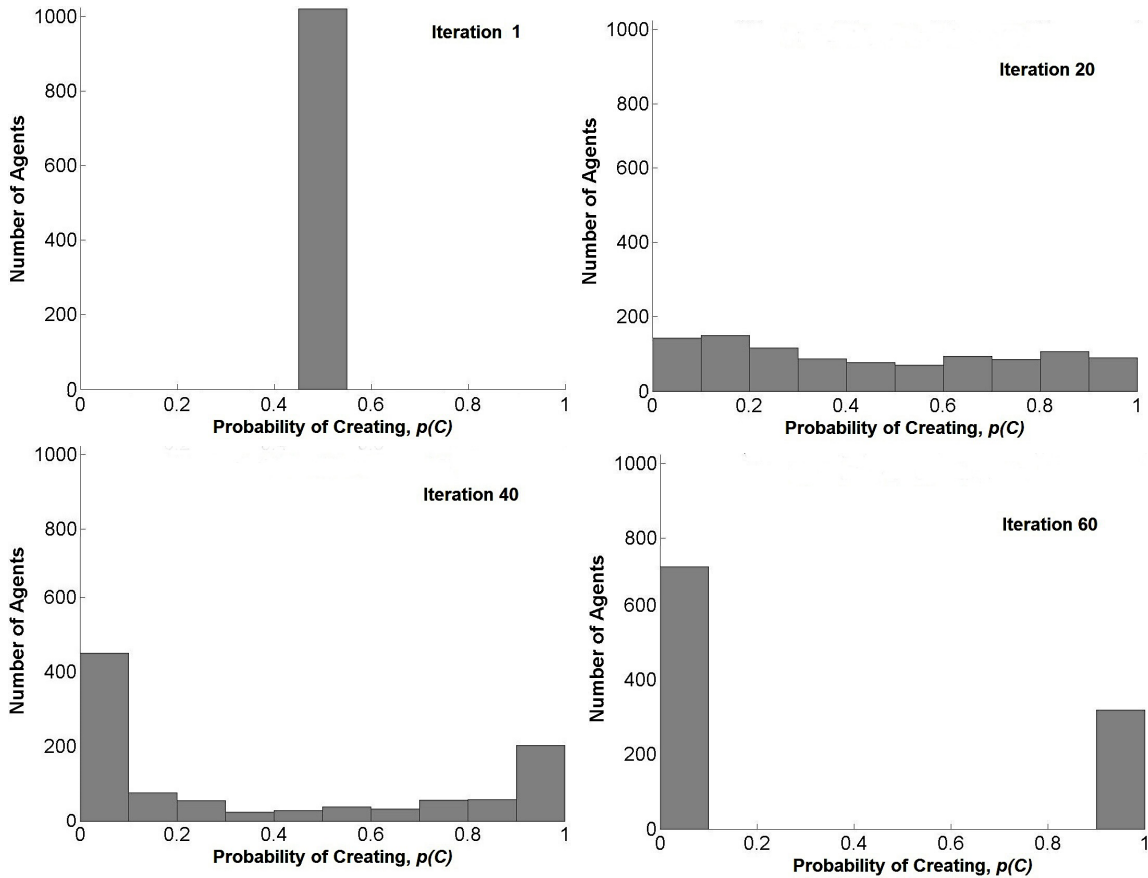

Figure 9. At the beginning of the run all agents created and imitated with equal probability. Midway through their *p(C)* values were distributed along the range from 0 to 1. By iteration 60 they had segregated into two distinct groups: conformers (with *p(C)* from 0 to 0.1) and creators (with *p(C)* from 0.9 to 1). (From Gabora & Tseng, 2014).

## Documenting Our Cultural Ancestry

The application of phylogenetic techniques derived from Darwinian approaches to culture present a distorted picture of cultural history as branching rather than network-like, because they do not incorporate horizontal transmission and blending. Also, because they incorporate only measurable attributes, they do not capture relatedness that resides at the conceptual level (e.g., mortars and pestles are highly related despite little similarity at the attribute level). To deal with this, the communal exchange theory of culture inspired a new technique for chronicling material cultural history, which has been used on multiple data sets. This method has been shown to generate a pattern of cultural ancestry that is more congruent with geographical distribution and temporal data than that obtained with phylogenetic approaches (Gabora, Leijnen, Veloz, & Lipo, 2011; Veloz, Tempkin, & Gabora, 2012). An example of a small set of data in **Figure 10** shows how the communal exchange representation is web-like—it allows for not just vertical but also horizontal lines of descent—not branching, as would be imposed on it by a phylogenetic representation, which allows for only vertical lines of descent.

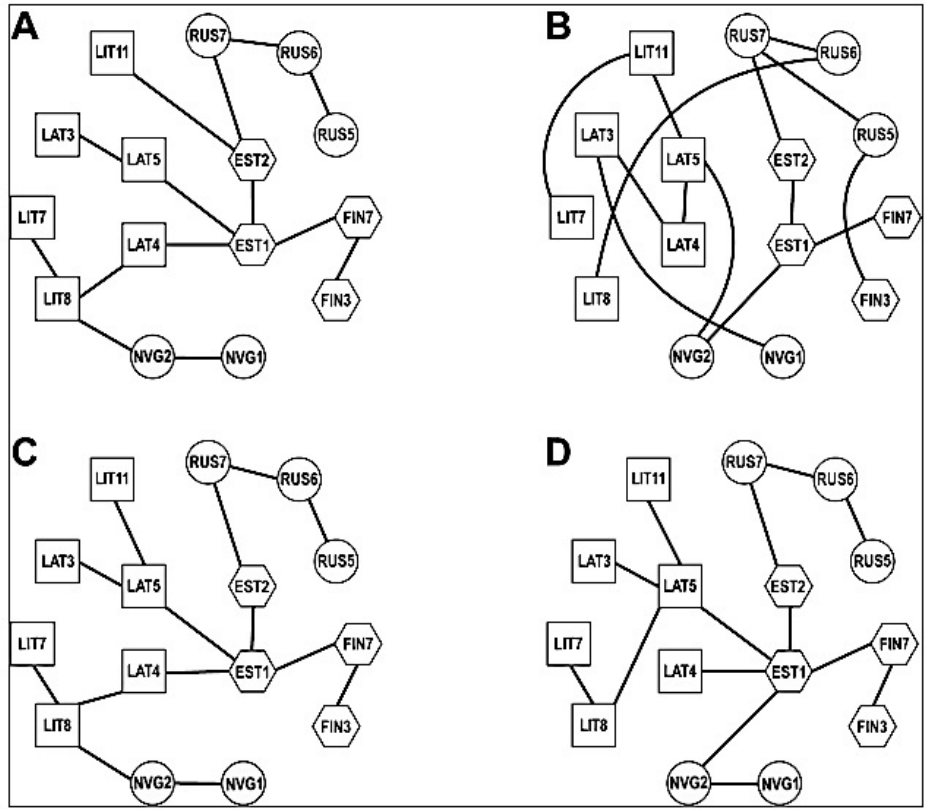

Figure 10. Similarity graphs based on conceptual network analysis of Baltic psalteries (a kind of stringed musical instrument) under different perspective weighting schemes. By incorporating not just physical characteristics but also conceptual attributes (such as those pertaining to sacred symbolic imagery), and weighting them differently, it was possible to resolve ambiguities obtained with the phylogenetic approach, and generated a lineage more consistent with other historical data. (A) Physical attributes only (B) Symbolism (C) Physical attributes and Symbolism (equal weights) (D) Physical attributes (25% weight) and Symbolism (75% weight). Each node corresponds to a single artifact. Node shapes indicate ethnolinguistic groups: Slavic (circle), Finnic (hexagon), and Baltic (square). Shaded nodes designate archaeological instruments (10-13 cc); remaining nodes correspond to ethnographical instruments (17-20 cc). (From Veloz, Tempkin, & Gabora, 2012).

## Understanding Creative Works in their Cultural Context

We now examine studies with real humans that explore in various ways the interplay between creativity and the cultural milieu.

### Cross-domain Recognizability of Creative 'Essence'

To what extent is a creative idea an abstract Platonic essence, and to what extent is it something that could only emerge from within the constraints of a specific domain? This question has implications for the evolution of culture, because the greater the extent to which ideas are not tied to their manifestation in any particular domain, the greater the extent to which creative novelty can reflect influences from diverse sources.

In a study that set out to investigate this empirically, it was demonstrated that when pieces of music were re-interpreted as paintings, naïve participants were able to correctly identify at significantly above chance which piece of music inspired which painting (Ranjan, Gabora, & O'Connor, 2014; Ranjan, 2014). Although the medium of expression is different, something of its essence remains sufficiently intact for an observer to detect a resemblance between the new work and the source that inspired it. This result lent empirical support to the largely anecdotal body of evidence that cross-domain influence is a genuine phenomenon, and suggested that, at their core, creative ideas are less domain-dependent than is generally assumed. It did not, however, provide evidence that the phenomenon extends beyond the artificial conditions of such a study, nor did it give an indication of how prevalent it is.

## Cross-Domain Influences on Innovation

Who or what influenced your creative output? Anecdotal reports and case studies (e.g., Feinstein, 2006) are one method of getting answers to this question. However, we are exposed to so many different people and objects in our lives that we may not be completely conscious of who or what influences our creative outputs. We are just beginning to be able to corroborate anecdotal reports with machine learning techniques designed to resolving lines of influence (Saleh, Abe, & Elgammal, 2014). These techniques are not yet able to discern cross-domain influences, wherein a creator in one domain (e.g., artist) is influenced by another domain (e.g., music).

Because Darwinian theories assume strictly vertical transmission and do not allow different "species" of cultural artifacts to "mate", they are incompatible with cross-domain influence. Communal exchange theory predicts that cross-domain influence is not just present but fuels cultural innovation. In a project designed to test between these predictions, 66 creative individuals in a variety of disciplines were asked to list as many influences on their creative work as they could (Gabora & Carbert, 2015). Results suggest that cross-domain influences are in fact more widespread than within-domain influences, even when broad within-domain influences (e.g., technology influenced by music) as well as narrow within-domain influences (e.g., music influenced by other music) are taken into account.

Table 2: Percentage of cross-domain (CD), within-domain narrow (WD-n), within-domain broad (WD-b), and uncertain (U) influences. From (Gabora & Carbert, 2015).

|            | CD  | WD-n | WD-b | U   |
| ---------- | --- | ---- | ---- | --- |
| % of Total | 47% | 27%  | 8%   | 18% |

## Cross-domain Recognizability of Creative Style

If creative output reflects not just chance or expertise but the idiosyncratic process of wrestling with personally meaningful issues to forge an integrated worldview, one might expect that creative individuals may have a characteristic style or 'voice', a distinctive facet of their personality that is recognizable not just within domains but across domains. Empirical support has been obtained for this prediction (Gabora, 2010b; Gabora, O'Connor, & Ranjan, 2012; Ranjan, 2014). Art students were able to identify at significantly above-chance levels which famous artists created pieces of art they had not seen before. They also identified at significantly above-chance levels which of their classmates created pieces of art they had not seen before. More surprisingly, art students also identified the creators of non-painting artworks that they had not seen before.

Similarly, creative writing students were able to identify at significantly above-chance levels passages of text written by famous writers that they had not encountered before and passages of text written by their classmates that they had not encountered before. Perhaps most surprising of all, creative writing students also identified at significantly above-chance levels which of their classmates created a work of art, that is, a creative work in a domain other than writing.

Creativity is sometimes thought to be "domain-specific" because expertise or eminence in one creative endeavor is rarely associated with expertise or eminence in another (Baer, 1996; Tardif & Sternberg, 1988). Although polymaths exist, famously creative scientists are rarely famously creative artists. The results of these studies support the view that creative achievement can be characterized in terms of, not just expertise or eminence, but the ability to express what we genuinely are through whatever media we have at our disposal. When looked at this way, creativity does appear to be more domain-general than it has been thought to be.

## Applications of Research on the Cultural Evolution of Creativity

Finally, let us briefly mention how research on creativity in a cultural context is being used to model one kind of cognitive process by which people find creative solutions to practical problems: *exaptation*. The concept of exaptation comes from biology where it refers to the situation wherein a trait that originally came about to solve one problem is co-opted for another use. Exaptation has been shown to play a pivotal role in economics (Dew, Sarasvathy, & Ventakaraman, 2004), and a preliminary attempt has been made to develop a mathematical model of exaptation that can be applied across disciplines (Gabora, Scott, & Kauffman, 2013). Applied to culture it refers to the situation wherein a different context suggests a new use for an existing item. The approach provides a formal model of what Rothberg (2015) calls Janusian thinking, which involves achieving a creative outcome by looking at something from a different perspective. Waste recycling is an interesting form of cross-domain creative influence because of its applications to sustainability efforts. An item that is a wasteful byproduct in one context is found to be useful in a different context. The approach has been used to model the creative restructuring of a concept in a new context when it is considered from another perspective (Gabora & Carbert, 2015). A similar approach can be taken to data transformation, in which data in one format is changed to a different format while preserving the content so that it can be put to a different purpose or made easier to interpret or understand (as in the visualization of astronomical data).

## Conclusions

To understand how culture evolves one must understand the creative processes that fuel cultural innovation. This chapter provided an overview of ways in which creativity and cultural evolution interact, and ways in which theoretical investigations and empirical studies of both can mutually inform one another. The chapter merely touches the tip of the iceberg of this fascinating topic. It will be exciting to see what the coming decade brings as we move ahead on the exciting journey toward understanding the lineages of creative influence that result in not just the artistic masterpieces that inspire us and the technological achievements that connect us, but the ideas, gadgets, and jokes we encounter each day that make life simpler or just make us smile.

**Acknowledgements**

This research was supported in part by a grant from the Natural Sciences and Engineering Research Council of Canada. Thanks to Simon Tseng, Stefan Leijnen, and Hadi Firouzi for the contributions to EVOC.